\documentclass[fleqn,usenatbib]{mnras}

\usepackage{newtxtext,newtxmath}
\usepackage{booktabs}

\usepackage[T1]{fontenc}
\usepackage{ae,aecompl}

\usepackage{graphicx}	
\usepackage{amsmath}	
\usepackage[capitalise]{cleveref}
\usepackage{hyperref}
\hypersetup{
    colorlinks = true
}

\usepackage{subcaption}
\usepackage{xcolor}

\newcommand{\Sersic}{S\'ersic~}

\title[Quiescent cluster galaxies at $z\sim2$]{Quiescent galaxies in a virialized cluster at redshift 2: Evidence for accelerated size-growth}

\author[E. Noordeh et al.]{E. Noordeh$^{1,2}$\thanks{E-mail: emiln@stanford.edu},
R. E. A. Canning$^{3}$,
J. P. Willis$^{4}$,
S. W. Allen$^{1,2,5}$,
A. Mantz$^{1,2}$, \newauthor
S. A. Stanford$^{6}$,
G. Brammer$^{7,8}$
\\
$^{1}$Department of Physics, Stanford University, 382 Via Pueblo Mall, Stanford, CA 94305-4060, USA\\
$^{2}$Kavli Institute for Particle Astrophysics and Cosmology, Stanford University, 452 Lomita Mall, Stanford, CA 94305-4085, USA\\
$^{3}$Institute of Cosmology and Gravitation, University of Portsmouth, Portsmouth, PO1 3FX, UK\\
$^{4}$Department of Physics and Astronomy, University of Victoria, 3800 Finnerty Road, Victoria, V8P 5C2 BC, Canada\\
$^{5}$SLAC National Accelerator Laboratory, 2575 Sand Hill Road, Menlo Park, CA 94025, USA\\
$^{6}$Department of Physics, University of California, One Shields Avenue, Davis, CA 95616, USA \\
$^{7}$Cosmic Dawn Center (DAWN), Jagtvej 128, DK2200 Copenhagen N, Denmark\\
$^{8}$Niels Bohr Institute, University of Copenhagen, Lyngbyvej 2, DK2100 Copenhagen \O, Denmark
}

\date{Accepted XXX. Received YYY; in original form ZZZ}

\pubyear{2021}

\begin{document}
\label{firstpage}
\pagerange{\pageref{firstpage}--\pageref{lastpage}}
\maketitle

\begin{abstract}
We present an analysis of the galaxy population in XLSSC 122, an X-ray selected, virialized cluster at redshift $z=1.98$. We utilize HST WFC3 photometry to characterize the activity and morphology of spectroscopically confirmed cluster members.
The quiescent fraction is found to be $88^{+4}_{-20}$ per cent within 0.5$r_{500}$, significantly enhanced over the field value of $20^{+2}_{-2}$ per cent at $z\sim2$.
We find an excess of ``bulge-like'' quiescent cluster members with \Sersic index $n>2$ relative to the field.
These galaxies are found to be larger than their field counterparts at 99.6 per cent confidence, being on average $63^{+31}_{-24}$ per cent larger at a fixed mass of $M_\star = 5\times10^{10} M_\odot$.
This suggests that these cluster member galaxies have experienced an accelerated size evolution relative to the field at $z>2$.
We discuss minor mergers as a possible mechanism underlying this disproportionate size growth.
\end{abstract}

\begin{keywords}
galaxies: clusters: general -- galaxies: evolution -- galaxies: high-redshift -- galaxies: structure -- galaxies: stellar content -- galaxies: fundamental parameters
\end{keywords}



\section{Introduction}

The nature of galaxy evolution from the early universe to the present day is intricately linked to large-scale environment. Denser environments tend to host more massive galaxies \citep[e.g.][]{Hogg2003,Baldry2006} that have had an accelerated evolution towards quiescence and bulge-dominated morphologies \citep[e.g.][]{Dressler1980,Butcher1978}.
This is exceptionally clear in the nearby universe where galaxy clusters are host to the most massive quiescent galaxies containing uniformly old populations of stars \citep[e.g.][]{Kodama1997,Mei2009}.
It is thought that this early-type cluster galaxy population is in place by $z\sim1$ \citep[e.g.][]{Lidman2008} with the bulk of star formation having occurred in these galaxies at $z\gtrsim2$ \citep[e.g.][]{Strazzullo2006,Andreon2013,Newman2014}.

The extreme overdensities of galaxy clusters can influence the evolution of their member galaxies through a variety of processes such as ram pressures stripping \citep[e.g.][]{Gunn1972,Ebeling2014}, tidal interactions \citep[e.g.][]{Farouki1981,Moore1996,Moore1999}, and ``strangulation'' \citep[e.g.][]{Larson1980,Bekki2002}. 
Galaxy clusters also play an important role in the evolution of their member galaxies by regulating the frequency of galaxy mergers.
During the formation of clusters, when galaxy densities are high and velocity dispersions are low, galaxy interaction rates are enhanced relative to the field \citep[e.g.][]{McIntosh2008,Lin2010,Kampczyk2013,Delaye2014}. However, the subsequent high-velocity dispersion of massive clusters greatly inhibits the rate of future mergers among its members \citep[e.g.][]{Mamon1992}.
This merger-driven dynamic has been suggested to significantly influence galaxy morphologies \citep[e.g.][]{Bekki1998,Matharu2019}, star-formation activity \citep[e.g.][]{Brodwin2013}, and active galactic nuclei \citep[e.g.][]{Ehlert2014,Noordeh2020}. 

While galaxy clusters serve as important laboratories to probe galaxy evolution and investigate the relative importance of environmental and secular processes, studies become increasingly rare at high redshifts. This is due to both a decline in the number of massive clusters at high redshift as well as increasing observational difficulty in reliably identifying them.
In particular, at high redshift, clusters are increasingly selected in the IR through their member galaxy population, which introduces biases into galaxy population studies.
Rather, these clusters should ideally be identified through an observational proxy that is less dependent on the properties of the galaxy population, such as the Sunyaev-Zeldovich effect or through X-ray emission of the Intra-Cluster medium (ICM). 
While the X-ray emitting ICM may be partially composed of stripped/ejected material from galaxies prior to the observing epoch, it is not directly dependent on the current state of star-formation in the member galaxies; whereas IR color, red-sequence cluster selection techniques are.
To date, only two clusters at $z\gtrsim1.8$ with extended X-ray emission have been reported \citep{Gobat2011,Newman2014}, however both clusters were initially selected in the IR through their member galaxy population.

In this paper we examine the cluster member population of XLSSC 122, an ICM selected, mature cluster at $z=1.98$ with 37 spectroscopically confirmed members \citep{Willis2020}.
XLSSC 122 is the highest redshift, ICM selected cluster discovered to date and provides a unique opportunity to investigate the influence of environment on galaxy evolution in the regime where galaxy growth is expected to be most rapid.

In \cref{sec:data} we describe our observations and detail our methodology for characterizing galaxy quiescence, mass, and size in both XLSSC 122 and a CANDELS control field. We investigate the distribution and structure of the passive and active cluster member population
in \cref{sec:galpops}. In \cref{sec:masssize} we compute the mass-size relation of the quiescent cluster members and compare to that of the field at $z\sim2$. Finally, we summarise our findings in \cref{sec:conclusion}.

\indent All magnitudes quoted in this work are AB magnitudes. Distances are computed adopting a cosmology with $\Omega_{M}=0.3$, $\Omega_\Lambda = 0.7$, and $H_0=70~\mathrm{km~s^{-1}~Mpc^{-1}}$. Uncertainties are quoted at the 68 per cent confidence level following \cite{Cameron2011} for Binomial population proportions. Cluster radii are measured in units of $r_{500}$, which is defined as the radius within which the mean density of the cluster is 500 times the critical density at that redshift. Cluster masses are quoted as $M_{500}$ values, with these being the mass contained within a sphere of radius $r_{500}$.

\section{Data and Methods} \label{sec:data}

\subsection{XLSSC 122} \label{sec:xlssc122}

The galaxy cluster XLSSC 122 (XLSSU J021744.1$-$034536) was originally discovered through its ICM emission in the XMM Large Scale Structure survey \citep{Pierre2006,Willis2013}.
With 100 ks of follow-up XMM observations, it was found to have a temperature of $kT=5.0\pm0.7$\,keV, an emission-weighted metallicity of $Z/Z_\odot=0.33^{+0.19}_{-0.17}$, and cluster mass of $M_{500}=(6.3\pm1.5)\times10^{13}M_\odot$ \citep{Mantz2014,Mantz2018}. 

This study utilizes Hubble Space Telescope (HST) Wide Field Camera 3 (WFC3) imaging of XLSSC 122 in the F105W and F140W bands as well as slitless spectroscopy with the G141 grism. The observations extend out to $2r_{500}$, approximately the virial radius of the cluster. The same data were previously used in \cite{Willis2020} to confirm XLSSC 122 as a mature cluster at $z_{\mathrm{cluster}}=1.98\pm0.01$ and robustly identify 37 member galaxies.

\subsection{Photometry, spectroscopy, and member selection} \label{sec:selection}

The photometric and spectroscopic data reduction and processing procedures used are described in detail in \cite{Willis2020}. In short, the F105W and F140W imaging were reduced using \verb+Grizli+ \citep{Grizli} and processed with \verb+SExtractor+ \citep{SExtractor} to produce source catalogs with AB magnitudes measured within 0.8 arcsecond circular apertures. The F140W segmentation map was used to identify undispersed source positions which we employed to build a full field spectral contamination model for each G141 image. With G141 observations split into four orientations, we were able to build robust contamination models for most sources in the crowded field and extract two-dimensional spectra for each source.

These spectra were cross-correlated with a suite of galaxy templates over the redshift range $0.2<z<4$ from which a probability distribution function (PDF) for the redshift of each source were derived. Cluster membership was split into two categories: ``gold'' and ``silver'', representing high and moderate probability members respectively. In this study we focus solely on gold members with F140W~$<24$, $r<2r_{500}$, and $P_{\mathrm{mem}}>0.5$. $P_{\mathrm{mem}}$ is the integral of the redshift PDF over $1.96<z<2.00$, an interval corresponding to $z_{\mathrm{cluster}}\pm3\sigma_z$, where $\sigma_z$ is the expected observed frame velocity dispersion of a 5-keV galaxy cluster \citep{Willis2020}. There are 28 such members identified in XLSSC 122.

\begin{figure}
 \includegraphics[width=0.47\textwidth]{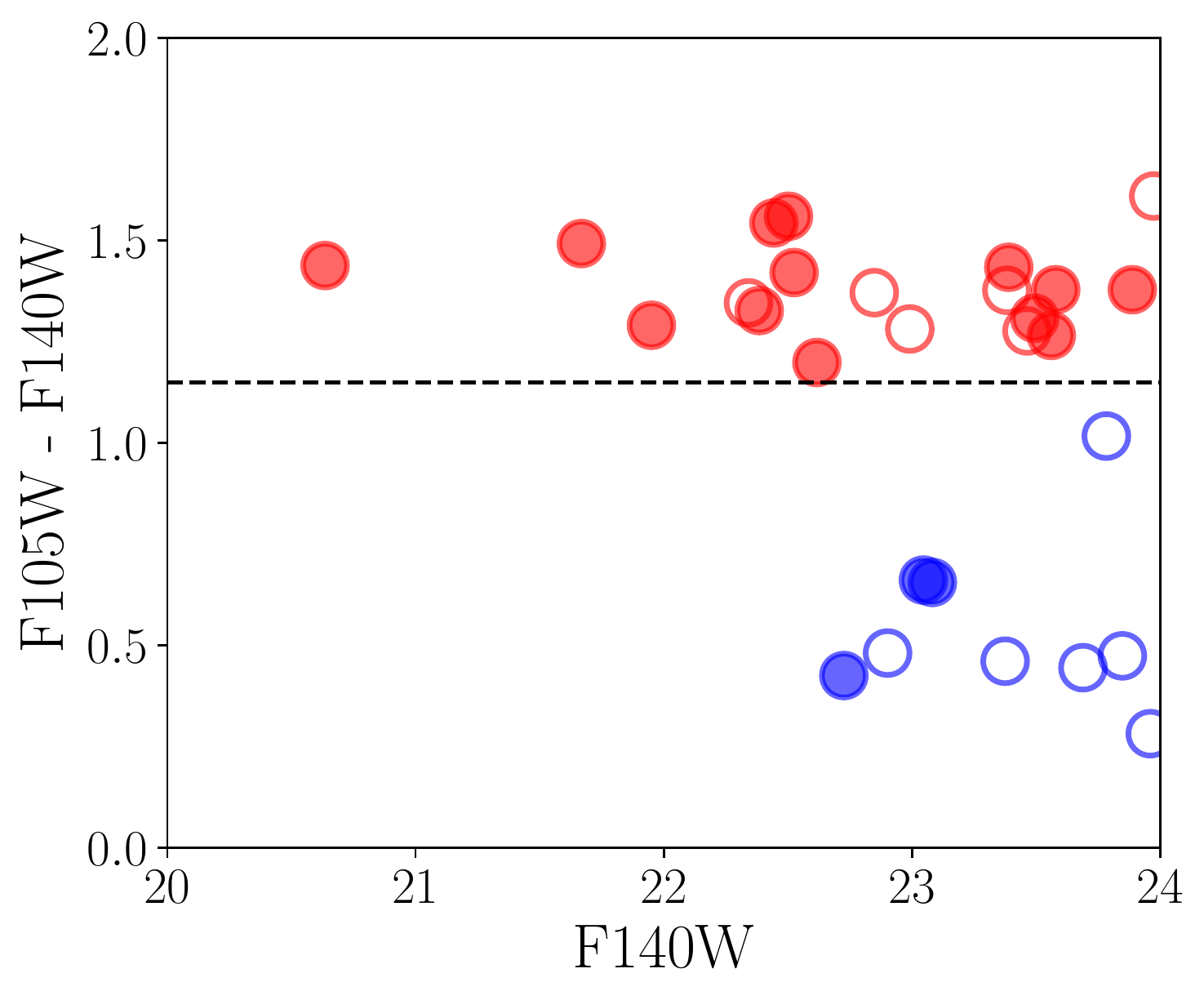}
 \caption{
 The color-magnitude diagram of XLSSC 122 is shown for spectroscopically confirmed cluster members within $2r_{500}$. The dashed line at $\mathrm{F105W}-\mathrm{F140W}=1.15$ demarks the separation between galaxies classified as quiescent, marked in red, and those classified as star forming, marked in blue. Galaxies with \Sersic index $n<2$ are marked by open circles while those with $n>2$ are marked as filled circles.
 }
 \label{fig:CMD}
\end{figure} 

\subsection{Galactic activity} \label{sec:activity}

We classify XLSSC 122 member galaxies as quiescent based on the red-sequence of the cluster color-magnitude diagram (CMD) shown in \cref{fig:CMD}. Sources with F105W-F140W $> 1.15$ are classified as quiescent whereas bluer sources are considered star-forming.
Color-magnitude selection is a good proxy for galaxy specific star formation rates (sSFRs) as illustrated in \cref{fig:field_cmd}.

\subsection{Field comparison sample} \label{sec:field}

We establish a field comparison sample by utilizing the AEGIS, COSMOS, GOODS-S, and UDS fields from CANDELS/3D-HST
\citep{Koekemoer2011,Brammer2012,Skelton2014,Momcheva2016}. We select sources in the redshift range $1.9<z<2.1$ using spectroscopic redshifts when available and photometric redshifts otherwise \citep[$z_{\mathrm{best}}$ from ][]{Momcheva2016}. We select only galaxies with F140W $<24$ to match our cluster sample and remove sources with bad photometry flags from \cite{Skelton2014}. There are a total of 417 field galaxies satisfying these criteria.

\begin{figure}
 \includegraphics[width=\columnwidth]{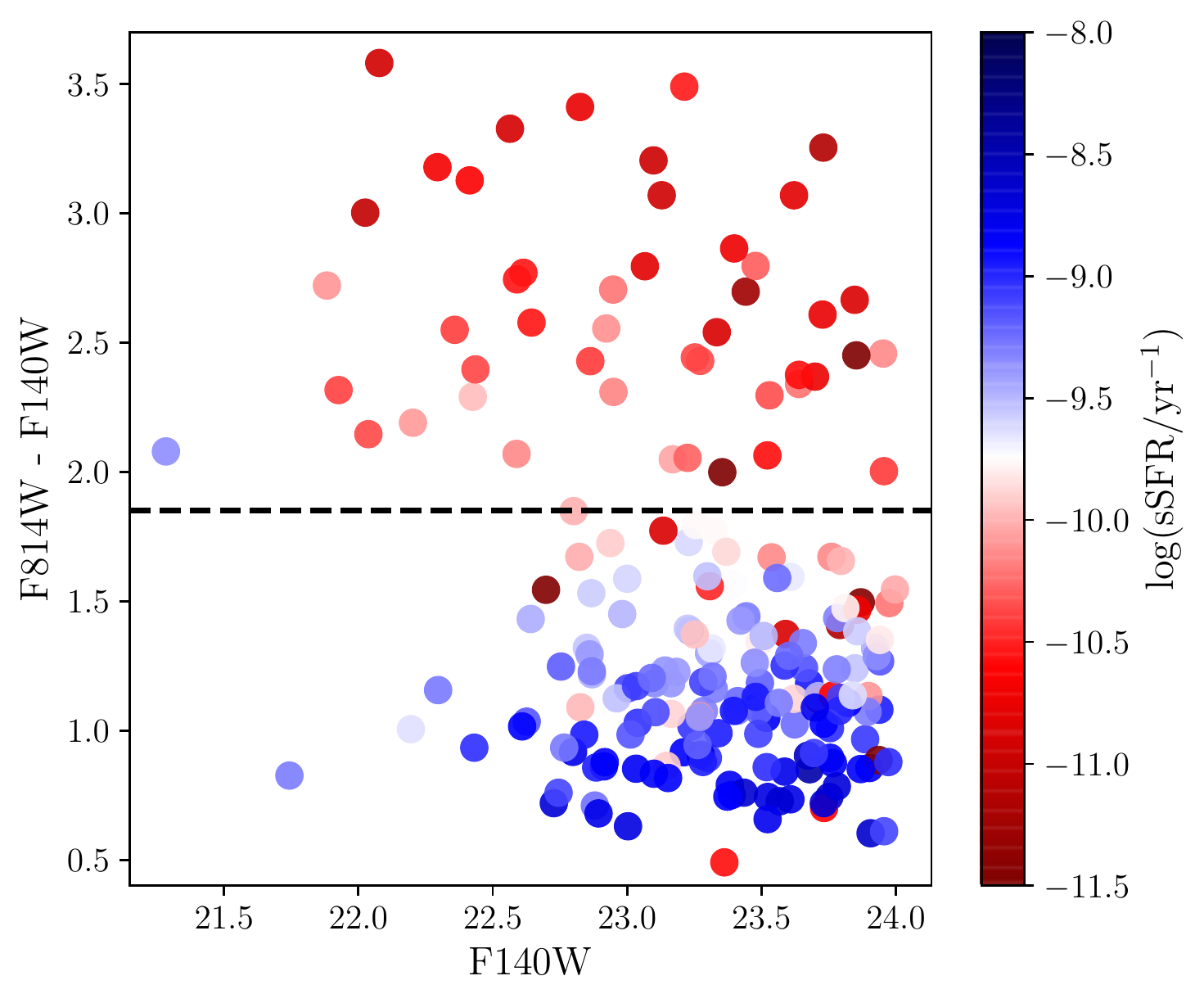}
 \caption{$\mathrm{F814W}-\mathrm{F140W}$ vs. $\mathrm{F140W}$ color-magnitude diagram of the CANDELS/3D-HST fields. Galaxies are selected as those with $1.9<z<2.1$ and F140W~$<24$. Only galaxies that have cataloged star-formation rates from \protect\cite{Whitaker2014} are shown. A clear trend in specific star formation rate ($\mathrm{sSFR}$) is observed wherein, on average, redder galaxies have substantially lower $\mathrm{sSFRs}$. In this study, we identify all field galaxies with $\mathrm{F814W}-\mathrm{F140W}>1.85$ as quiescent and those with $\mathrm{F814W}-\mathrm{F140W}<1.85$ as star forming.}
 \label{fig:field_cmd}
\end{figure}

Rather than selecting quiescent field galaxies based on their cataloged star-formation rates \citep{Kriek2009,Skelton2014,Whitaker2014}, we attempt to make the quiescent galaxy selection as similar to that of XLSSC 122 as possible. While the CANDELS/3D-HST fields have F140W imaging, there is no F105W data available. Instead, we utilize the F814W filter (the closest available filter to F105W) to construct the F814W-F140W vs. F140W CMD shown in \cref{fig:field_cmd}.
We visually identify the red sequence in this CMD as F814W-F140W~$>1.85$ and use this threshold to classify quiescent galaxies comparable to those selected in XLSSC 122\footnote{We use the PySynphot package \citep{synphot} to confirm that this threshold separates comparable $z=2$ galaxies as the F105W-F140W threshold described in \cref{sec:activity}. This is done by first finding all $z=2$ stellar templates from \cite{Bruzual2003} that would give the same galaxy colors as the quiescent and star forming XLSSC 122 members shown in \cref{fig:CMD}. We then simulate the F814W-F140W colors of these templates and find that this threshold cleanly separates the same quiescent and star-forming galaxies.}.
We identify 84 quiescent field galaxies following this procedure. We perform our analysis adjusting this threshold by $\pm0.2$ and find no significant change in our results.

\subsection{Sizes} \label{sec:galfit}

We fit all galaxies in XLSSC 122 and our field comparison sample with a single component, two-dimensional \Sersic profile. This profile takes the functional form:

\begin{equation}
    I(r) = I(r_e)\exp\Bigg\{-\kappa\Bigg[\bigg(\frac{r}{r_e}\bigg)^{1/n}-1\Bigg]\Bigg\} ,
\end{equation}

\begin{equation}
    r = \Bigg[ (x-x_c)^2 + \bigg( \frac{y-y_c}{q}\bigg)^2 \Bigg]^{1/2} .
\end{equation}

\noindent Here $r$ is the radial distance of pixel location $(x,y)$ from the source center at $(x_c,y_c)$, where the coordinate axes are aligned with the principle axes of the ellipse. $I(r)$ is the intensity at radius $r$, $q$ is the axis ratio, $n$ is the \Sersic index, $r_e$ is the half-light radius, and $\kappa$ is an $n$ dependent normalization constant. \footnote{Following \cite{VanderWel2014,Newman2014,Matharu2019}, the effective radii used in our study are not circularized as is sometimes done in the literature. However, for comparison to such studies, we perform our analysis with circularized radii as well ($r_e^{\mathrm{circ}}=r_e\sqrt{q}$) and find no significant change to our conclusions. For reference, the circularized radii of cluster members are provided in \cref{tab:members}.}

We perform these fits in the F140W band, for both the cluster and the field, using \verb+GALFIT+ \citep{Peng2002}, closely following the methodology of \cite{VanDerWel2012} and \cite{Matharu2019}. We start by making square image cutouts centered on each object with a width equal to 10 times the \verb+SExtractor+ determined half-light radius. These images are sky-subtracted and have units of electrons s$^{-1}$. We also produce a noise map for each fitting region by first computing the intrinsic variance for each pixel from the drizzled weight images.
To this we add the variance at each pixel from the Poisson noise due the sources themselves. We take the square root of this total variance and divide by the computed exposure time in each pixel to arrive at a noise map in electrons s$^{-1}$, matching our input images. This is provided as a ``sigma image'' to \verb+GALFIT+ for each source.

We simultaneously fit all neighbouring sources whose centers fall within 10 half-light radii of each object, provided they are no more than than 4 magnitudes fainter than the primary source. This allows us to account for any contamination from nearby sources.

\begin{figure}
  \includegraphics[width=\columnwidth]{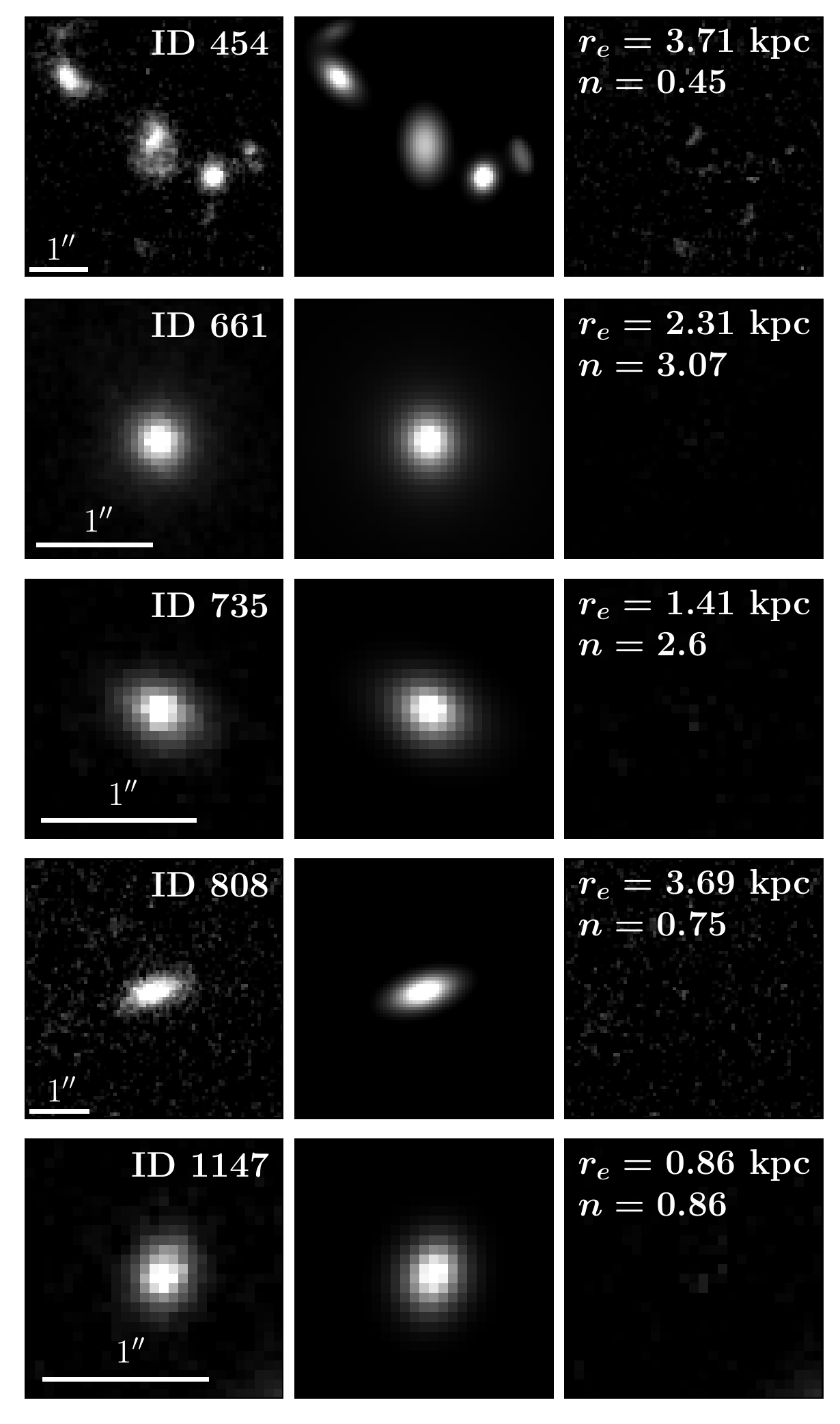}
  \vspace{-1.5em}
 \caption{\Sersic profile fits for a sample of XLSSC 122 members are shown. The first, second, and third columns show the F140W image, model fit, and residual cutouts for each source respectively. Square cutouts are centered on each source with a width equal to 10 times the SExtractor determined half-light radius as described in \cref{sec:galfit}. The source ID is shown in the first panel and the effective radius $r_e$, and \Sersic index, $n$, for the best fitting model is shown in the third panel.}
 \label{fig:galfit_sample}
\end{figure}

For the AEGIS, COSMOS, GOODS-S, and UDS
fields we utilize their respective published F140W Point Spread Functions (PSFs) from \cite{Skelton2014}. To construct the PSF of our XLSSC 122 galaxies we first identify ``clean'' stars in our WFC3 F140W imaging as those with F140W~$<20$ that have no contamination from nearby sources within a square region of width 3 arcseconds. We find three such stars with F140W magnitudes of 19.4, 19.7, 19.9. We then perform three separate \verb+GALFIT+ runs on the XLSSC 122 galaxies with the PSF set as: 1) the nearest clean star, 2) the brightest clean star, and 3) the published F140W PSF for the COSMOS field from \cite{Skelton2014}. For all sources but three, which are contaminated by significant Intra-Cluster Light (ICL; discussed below), we find that the measured $r_e$ agree within uncertainties for all three runs. This suggests that our choice of PSF is not impacting the accuracy of our measurements and we select the brightest clean star with F140W~$=19.4$ as our default PSF since it resulted in the highest precision measurements. This PSF is convolved with a $128\times128$ pixel region ($\sim8\times8$ arcseconds) centered on each model component during the fitting routine. A set of sample fits to XLSSC 122 members are shown in \cref{fig:galfit_sample}.

The Brightest Cluster Galaxy (BCG) in XLSSC 122 and two nearby gold quiescent members are contaminated by significant ICL and have large ($>20$ per cent) errors on the measured $r_e$ following the procedure above. These sources are refit with a larger fitting region of 40 times the half light radius which resulted in visually confirmed robust fits with significantly smaller errors. 

We identify failures in our fitting procedure as those with $\sigma_{r_e} / r_e > 0.3$. There are zero failures among XLSSC 122 members and 9 in the control field. We visually inspect these failures and find them to be almost entirely associated with low surface brightness, irregular galaxies that are poorly characterized by a single component \Sersic fit. These sources are excluded from our study going forward. 

We test our fitting procedure by refitting a $z\sim2$ sample of COSMOS galaxies and comparing to the published structural parameters in \cite{VanDerWel2012}. This comparison is detailed in \cref{app:test}. We find our measured $r_e$ to be on average 2.4 per cent smaller than the published values. This difference is reasonable given the average cataloged error of $6.4$ per cent on these high redshift sources and negligible for the purposes of our analysis.
The fitting region and convolution box-sizes described above have been tuned to maximize the agreement of our measured radii to the cataloged radii in this test.

\subsection{Stellar masses} \label{sec:masses}

We determine the stellar masses of the XLSSC 122 quiescent members by starting with the stellar masses published in \cite{Willis2020}, which were measured using F105W and F140W magnitudes from 0.8 arcsecond apertures. We apply a correction factor to these masses to account for the fact that fixed sized apertures will be missing light from larger sources and will be contaminated by the ICL for sources near the BCG. We do this by leveraging the fact that our \verb+GALFIT+ routine disentangles the source flux from the local ICL of each source by modelling the ICL as a background. We use our \Sersic fits to compute a fully integrated source flux and convert to a total mass value.
The mass correction factor, $\eta$, is computed as:
\begin{equation}
    m_F - m_G = 2.5 \log \eta ,
\end{equation}
where $m_F$ is the F140W magnitude from a fixed aperture and $m_G$ is the fully integrated magnitude from our best fitting \Sersic model. The value of $\eta$ ranges from $0.3$ for members significantly contaminated by ICL to $2.2$ for the largest cluster members. The stellar masses of the quiescent cluster members are cataloged in \cref{tab:members}.
Note that with only two bands of HST photometry, we are unable to reliably determine stellar masses for star-forming cluster members.

For the field comparison sample we utilize the cataloged stellar masses from \cite{Skelton2014} which are derived using the FAST code \cite{Kriek2009}. We correct these masses in the same way as the cluster members and find mass correction factors ranging from $0.8$ to $1.3$ with a mean of $1.04$.

\section{Passive and active galaxy populations} \label{sec:galpops}

\begin{figure}
    \includegraphics[width=0.99\columnwidth]{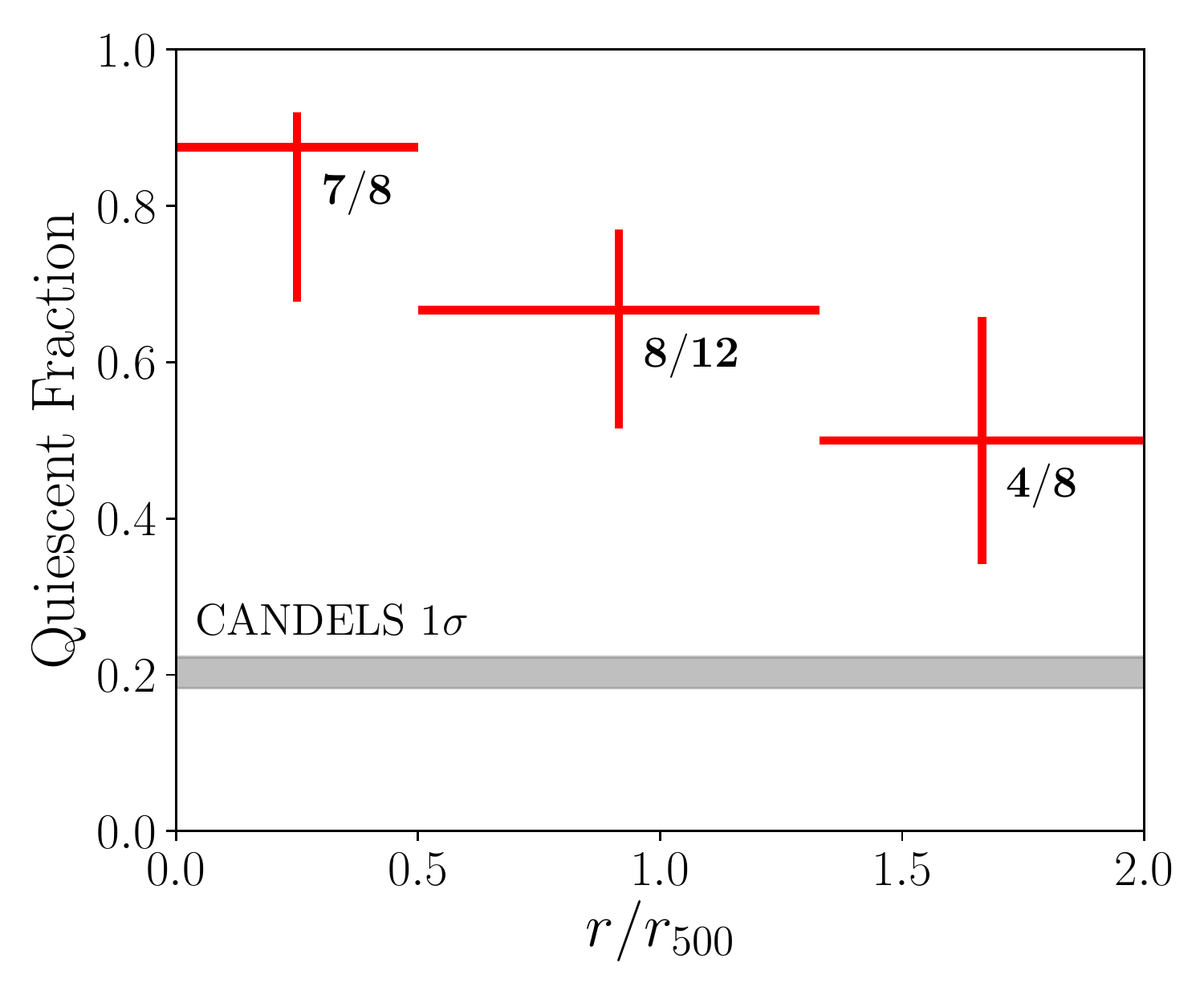}
 \caption{The quiescent fraction of spectroscopically confirmed XLSSC 122 members with F140W~$<24$ are shown in red as a function of cluster-centric radius. The number of quiescent galaxies out of the number of total galaxies is annotated at each data point. Errors indicate 1$\sigma$ binomial uncertainties. The gray band shows the 1$\sigma$ bounds of the quiescent fraction in the CANDELS field for galaxies with $1.9<z<2.1$ and F140W~$<24$.
 }
 \label{fig:QF}
\end{figure}

\subsection{Environmental quenching}
We show in \cref{fig:QF} the cluster quiescent fraction as a function of cluster-centric distance.
We find a quiescent fraction of  $88^{+4}_{-20}$ per cent within 0.5$r_{500}$, significantly enhanced over the field value of $20^{+2}_{-2}$ per cent.
Clearly, the cluster environment has played a powerful role in quenching its member systems, truncating star formation in its core in a remarkably similar fashion to massive clusters in the local Universe \citep[e.g.][]{vonderLinden2010,Mahajan2010}. Similar quiescent fractions have also been found in the cores of other high-$z$ clusters
\citep[e.g.][]{Strazzullo2013,Newman2014,Strazzullo2019}.
However, all such clusters at $z\gtrsim1.8$ were selected through the IR color of their galaxy populations, which can be biased towards selecting clusters with higher quiescent fractions.\footnote{Apart from XLSSC 122, the highest redshift ICM selected cluster discovered to date is SPT-CLJ0459-4947 at $z=1.72\pm0.02$ \citep{Strazzullo2019}}
For the first time, we are now observing the same behaviour in an ICM selected cluster at $z=1.98$, corresponding to a lookback time of more than 10 Gyrs.

XLSSC 122 is very unlikely to have assembled earlier than 1 Gyr prior to the epoch of observation \citep{Willis2020}, yet we observe a remarkably clear red sequence and a dramatically enhanced quiescent fraction in the cluster core (see \cref{fig:CMD,fig:QF}). This suggests that if environmental processes that operate on the halo mass scale of the cluster are responsible for quenching star formation, they must be relatively fast acting. Alternatively, processes that operate on smaller halo mass scales (e.g. in proto-cluster and group environments) and longer timescales, such as the pre-processing of cluster member galaxies through mergers, could be driving the truncation of star formation.

The continuous increase of the passive fraction from $2r_{500}$ to the core (\cref{fig:QF}) suggests ram pressure stripping or pre-processing by mergers may be the dominant transformation mechanisms acting on XLSSC 122 member galaxies \citep{Abadi1999, Fujita1998, Mihos1995}. Processes such as ``strangulation'',  ``starvation'', and ``harassment'' operate on the cluster mass scale over $\gtrsim$2 Gyr timescales \citep{McGee2009,Larson1980,Moore1999} and are therefore less likely. This is contrary to the conclusions of \cite{Treu2003} in their study of Cl 0024+16, a $z=0.4$ cluster, where they suggest harassment and starvation as the main drivers of the mild radial trend. They argue for a longer timescale due to the remarkable homogeneity of the red-sequence galaxy population in the cluster, which is also a characteristic of XLSSC 122.
However, \cite{Treu2003} notes that there is much scatter in the morphology-density relation in CL 0024+16 beyond the core and that many systems appear to retain a connection to their local substructures. As such, merging systems in pre-processed groups could be a common quenching mechanism in both systems.

\subsection{Galaxy morphologies} \label{sec:morphologies}

\begin{figure*}
    \centering
    \begin{subfigure}[b]{0.49\textwidth}
        \includegraphics[width=0.99\textwidth]{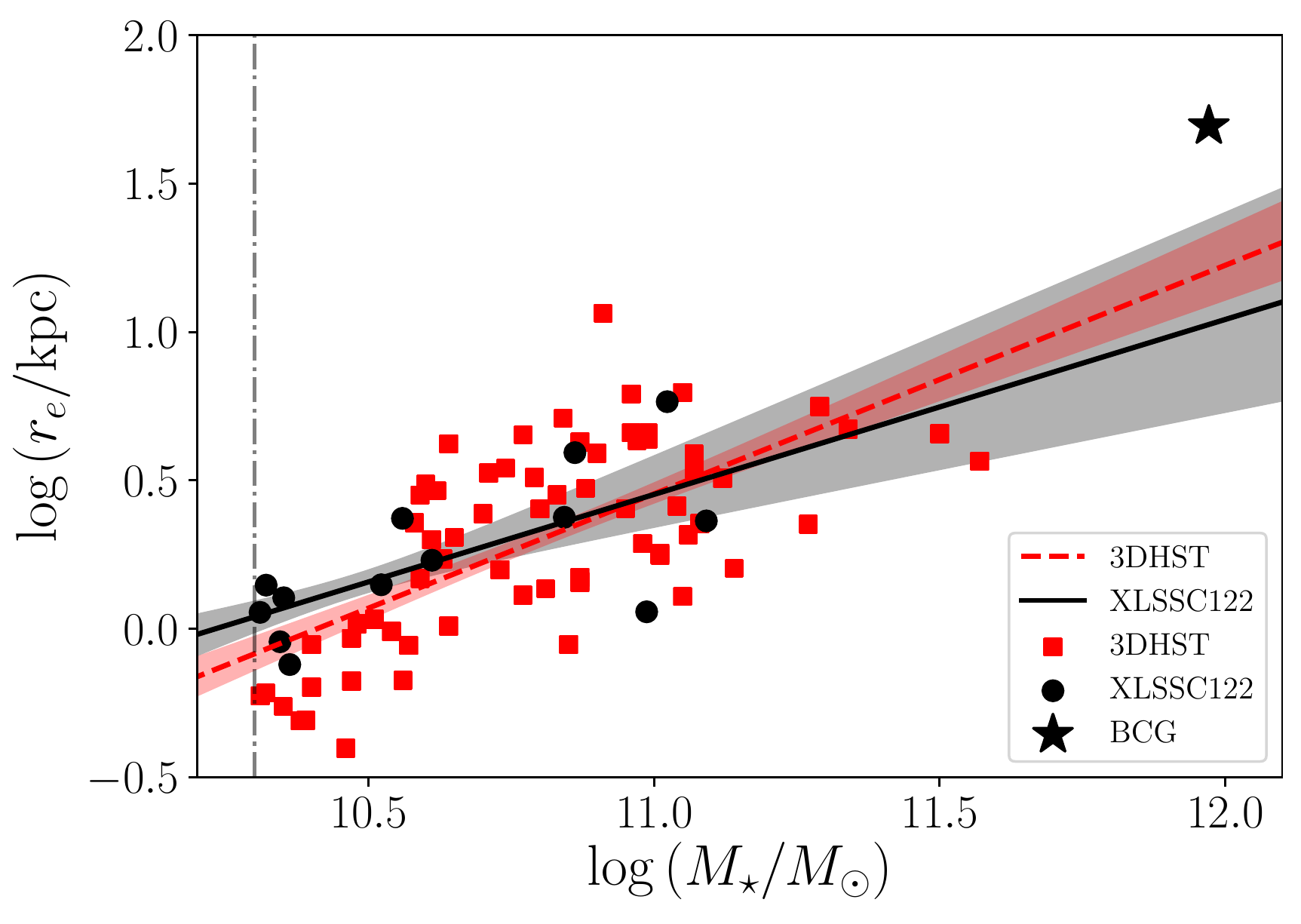} 
    \end{subfigure}
    \hfill
    \begin{subfigure}[b]{0.49\textwidth}
        \includegraphics[width=0.99\textwidth]{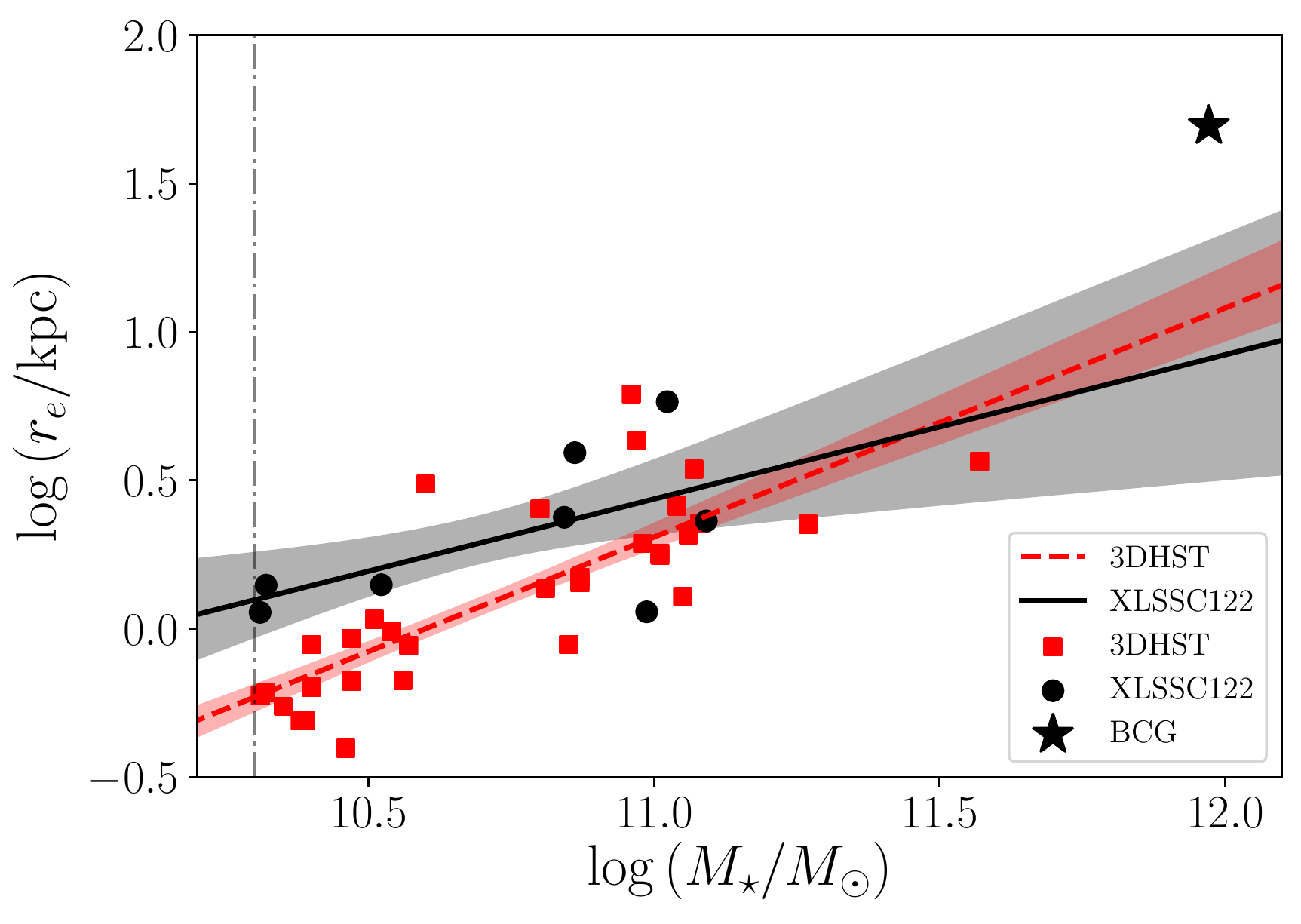}
    \end{subfigure}
    \caption{{\bf Left:} The effective radii and stellar masses of quiescent cluster and field galaxies are marked as black circles and red squares respectively. Solid black and dashed red lines indicate the best fitting mass-size relation for the cluster and field sample as detailed in \cref{sec:masssize}. The black and red contours outline the 1$\sigma$ uncertainties in the fit obtained through bootstrapping. The black star identifies the BCG of XLSSC 122, which is excluded when fitting for the cluster mass-size relation. The vertical dot-dashed line at $M_\star = 2\times10^{10} M_\odot$ marks the lower mass limit applied to this study. {\bf Right:} The same as on the left but for the bulge-like, $n>2$, quiescent population in both the cluster and field.}
    \label{fig:mass_size}
\end{figure*}

We investigate the morphological properties of the cluster members through their \Sersic indices as is commonly done in the literature \citep[e.g.][]{Strazzullo2013,Lani2013,Matharu2019}. We characterize galaxies with \Sersic indices $n<2$ and $n>2$ as ``disk-like'' and ``bulge-like'' respectively.

This morphological separation is illustrated on the cluster CMD in \cref{fig:CMD}. We find that $78^{+8}_{-19}$ per cent
of quiescent cluster members with $\log M_\star/M_\odot > 10.4$ are bulge-like, in agreement with the $70^{+10}_{-20}$ per cent quiescent fraction found by \cite{Strazzullo2013} for the same mass range in an IRAC selected cluster at $z=2$.
This is a significant enhancement relative to the $49^{+4}_{-5}$ per cent bulge-like fraction found for similarly massive quiescent galaxies in the CANDELS control field.
Our results are comparable to those found by cluster studies at low to intermediate redshifts \citep[e.g.][]{Moran2007,Cooper2012,Kuchner2017} where the cluster enhancement is typically attributed to environmental processes that cause a fading of the stellar disc relative to the inner bulge.

If it were environmental processes that were driving this morphological difference between XLSSC 122 and the field, we would expect to see some dependence on cluster-centric radius \citep[since environmental processes may depend strongly on local density and/or tidal forces, e.g.][]{Moran2007,Treu2003,Kuchner2017}. However, we find no physical segregation of galaxy morphologies in XLSSC 122, with sources characterized by a range of \Sersic indices existing across all cluster-centric radii. This contrasts with \cite{Strazzullo2013} where it was found that all $n>2$ systems fell within $\sim0.5r_{500}$ in an IRAC selected $z\sim2$ cluster.

Additionally, we find that only $33^{+18}_{-11}$ per cent of cluster star forming galaxies are bulge-like. This correlation between the star-formation and morphological properties of our cluster member galaxies is consistent with observations made in both high and low density environments at high-$z$ \citep[e.g.][]{Papovich2012,Patel2012}.

\section{Mass-size relation} \label{sec:masssize}

We compare the mass-size relation of quiescent XLSSC 122 members to those in the control field by fitting a relation of the form, 

\begin{equation} \label{eq:mass_size}
    \log \frac{r_e}{\mathrm{kpc}} = \alpha + \beta \log \frac{M_\star}{5\times10^{10}M_\odot}
\end{equation}

\noindent to both samples. Following \cite{VanderWel2014}, we use a mass pivot of $5\times10^{10}M_\odot$ in \cref{eq:mass_size} and we restrict our analysis to sources with $M_\star > 2\times10^{10} M_\odot$. Sizes for both samples are measured in the F140W band as described in \cref{sec:galfit}.
Our final sample consists of 13 quiescent cluster members and 68 field sources. 

We determine the best fitting model parameters and their uncertainties by performing a simple least-squares linear regression on 1000 bootstrapped samples of the cluster and field. The resultant mass-size relations are shown in the left panel of \cref{fig:mass_size}. We find no statistically significant difference between the two relations in these samples. Note that we exclude the BCG from the cluster fit as it has likely evolved through different processes than the general cluster galaxy population; however, our results do not significantly change with its inclusion.

\subsection{Morphological dependence} \label{sec:morph_fit}

We perform the same mass-size fits as above on the bulge-like, $n>2$, population in the cluster and the field. The results are shown in the right panel of \cref{fig:mass_size}. There are 8 such cluster galaxies and 29 field galaxies. Here we also exclude the BCG from the cluster fit but our results are independent of its inclusion.

\begin{figure}
 \includegraphics[width=0.46\textwidth]{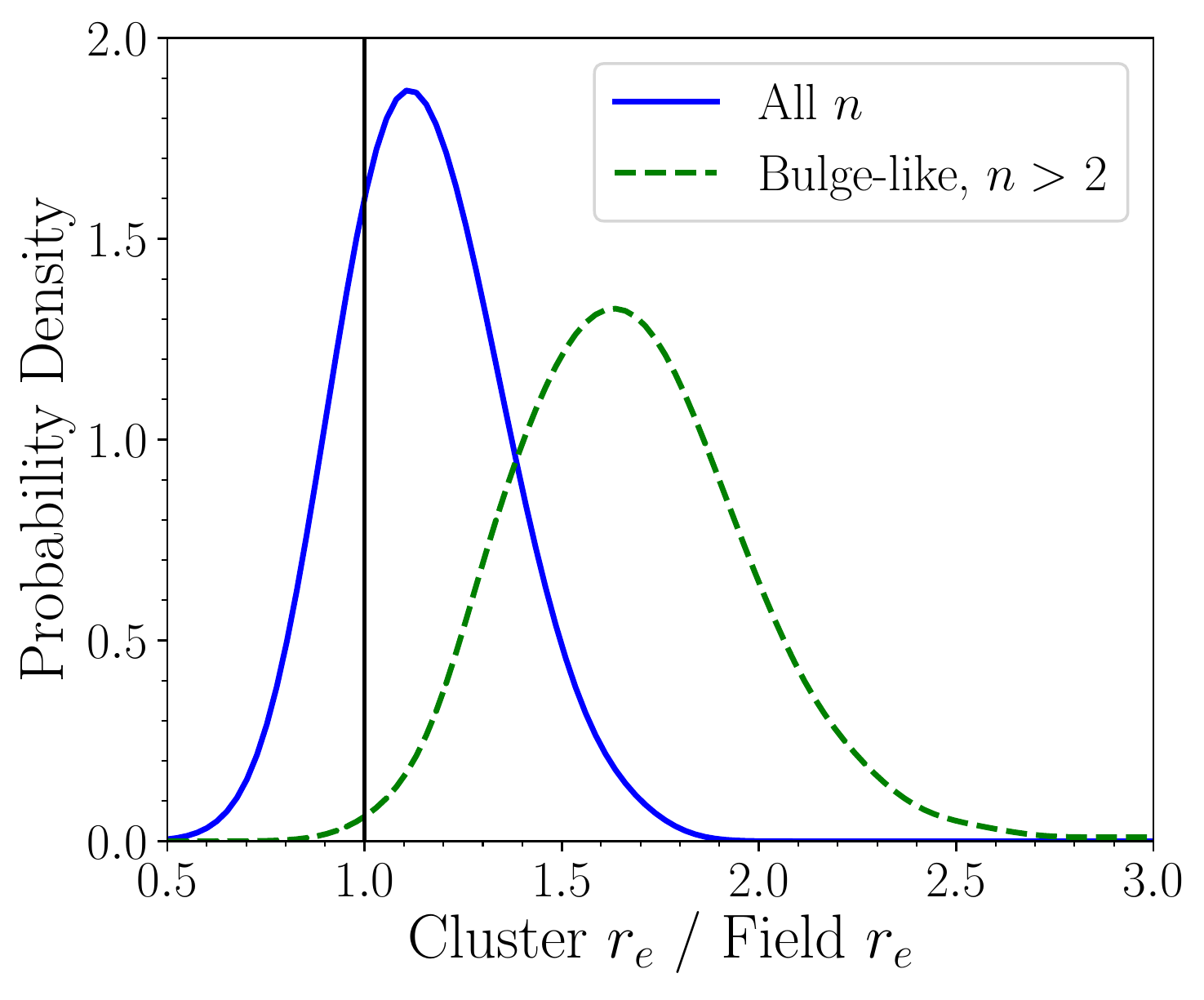}
 \caption{The probability densities for the ratio of the cluster to the field effective radius are shown for quiescent galaxies at a fixed mass of $M_\star = 5\times10^{10} M_\odot$. The result when no selection is made based on galaxy morphology is plotted in blue and is consistent with there being no difference between the field and cluster radii. The result when selecting only cluster and field quiescent galaxies with \Sersic index $n>2$ is shown in green, which suggests the cluster sizes are larger than the field at 99.6 per cent confidence.
 The probability densities are derived through bootstrapping as discussed in \cref{sec:masssize} and visualized with a Gaussian kernel density estimate with a bandwidth of 0.1.
 }
 \label{fig:mass_size_pdf}
\end{figure}

The posteriors in \cref{fig:mass_size_pdf} show the ratio between the cluster effective radius at a fixed mass of $M_\star=5\times10^{10}M_\odot$ (corresponding to $\alpha$ in \cref{eq:mass_size}) and that of the field. They are shown for all galaxy morphologies and only bulge-like morphologies. These distributions were derived by computing the ratio of $\alpha$ for the cluster and field fits across 1000 bootstrapped samples.

While there is no significant difference between galaxy sizes in the cluster and field when considering all morphologies, the result is different when looking at only bulge-like galaxies. These galaxies are found to be larger than their field counterparts at 99.6 per cent confidence, being on average $63^{+31}_{-24}$ per cent larger.

Our observations are broadly consistent with the limited number of previous studies performed at $z>1.6$, where quiescent cluster galaxies have also been found to be larger than their field counterparts \citep{Bassett2013,Zirm2012,Strazzullo2013,Papovich2012}. However, many of these studies look at proto-cluster environments and/or rely heavily on photometric redshifts. In none of these studies is the cluster ICM selected, which is crucial for galaxy population studies; this paper presents the first such data. Furthermore, due to small sample sizes and photometric redshift limitations, prior observations have only been able to claim larger sizes relative to the field at modest ($<2\sigma$) significance \citep[see][for a detailed comparison between the results of $z>1.6$ studies]{Matharu2019}.

At first glance our results may appear to contrast with those of \cite{Newman2014}, where no difference was found between quiescent galaxy sizes in the JKCS 041 cluster versus the field, even when selecting based on spheroid-dominated morphologies. At $z=1.8$, with 15 spectroscopically identified quiescent cluster members and diffuse ICM emission observed by {\it Chandra}, JKCS 041 is the most similar cluster to XLSSC 122 that has been studied. However, with a mass completeness limit of $\log M_\star/M_\odot > 10.6$, \cite{Newman2014} focus on higher mass galaxies than we do here. In particular, at a fixed mass of $M_\star=10^{11}M_\odot$, we find our bulge-like cluster galaxies to be larger than the field only at the $\sim1\sigma$ level, bringing our findings into agreement with those of \cite{Newman2014}.\footnote{\cite{Newman2014} characterizes the morphology of galaxies using the axis-ratio, $q$, of single \Sersic profile fits rather than the \Sersic index, $n$, used here. We test our results by applying an additional selection of $q>0.4$ and $q>0.5$ on our bulge-like sample and find no statistically significant difference in our fits.} Conversely, the difference becomes more significant for galaxies with $M_\star<5\times10^{10}M_\odot$. This same trend, where the difference between cluster and field galaxy sizes becomes more pronounced at smaller masses, was also found by \cite{Papovich2012} in a $z=1.62$ protocluster.

\subsection{Minor merger driven size growth} \label{sec:mergers}

Observations of quiescent galaxies in the field have almost unanimously observed galaxies growing larger in size over cosmic time, at fixed stellar mass \citep[e.g][]{Newman2012,VanderWel2014}. While the physical mechanisms behind this disproportionate growth in size relative to mass are only partially understood, minor mergers are commonly invoked as a plausible explanation \citep[e.g.][]{Trujillo2011,Ferreras2014,Matharu2019}.

In this study we observe a statistically significant enhancement of bulge-like quiescent cluster member sizes relative to the field. The morphological dependence of the mass-size relation that we observe can be understood in the context of minor merger driven size growth. Galaxies that have indeed been subject to enhanced merger histories are less likely to have maintained their fragile stellar disks and are expected to have more bulge-dominant light profiles \citep[e.g.][]{Hopkins2009b,Kormendy2009}.

At $z>1.3$, comparative studies between quiescent galaxies in cluster/proto-cluster versus field environments have typically found larger size galaxies to be preferentially located in denser environments \citep[e.g.][]{Papovich2012,Strazzullo2013,Delaye2014,Andreon2018}. The physical driver of these observations has also typically been attributed to merger driven size growth. Cluster member galaxies at $z\gtrsim1.3$ have likely experienced more mergers than co-eval field galaxies over their lifetime, driven by enhanced galaxy interaction rates during the formation of clusters when velocity dispersions are low and galaxy densities are high \citep[e.g.][]{Delaye2014}. If it is these interactions that are driving the disproportionate growth of galaxy sizes relative to mass, we would expect to see the enhanced cluster member sizes relative to the field that are observed.

Observations of the local universe ($z\lesssim0.2$) find no dependence of the quiescent mass-size relation on local density \citep[e.g.][]{Weinmann2009,Maltby2010,Cappellari2013}.
Therefore, there must be in place a mechanism for the field galaxies to catch-up in size with cluster members, or, conversely, for the growth of cluster galaxies to slow down relative to the field from $z\gtrsim1.3$ to the present. This is precisely what happens in the high-velocity dispersion environments associated with high-mass, virialized clusters, which greatly suppress the rate of interactions between cluster members \citep[e.g.][]{McIntosh2008,Mamon1992}. Thus, while initially subject to enhanced interaction rates in the proto-cluster/group environments preceding cluster formation \citep[e.g.][]{Papovich2012,Bassett2013,Lotz2013}, once accreted into a massive, virialized cluster, the future size-growth of cluster members is inhibited \citep[e.g.][]{Matharu2019}.

\section{Summary and conclusions} \label{sec:conclusion}

This study has analyzed the galaxy population in XLSSC 122, an ICM selected, virialized cluster at $z=1.98$. The relationship between star formation activity, stellar mass, and galaxy struture has been investigated for 28 spectroscopically confirmed cluster members with F140W~$<24$ and $r<2r_{500}$. We compare these galaxies to a similarly selected control field from CANDELS with $1.9<z<2.1$ and F140W~$<24$. Our primary findings are as follows:

\begin{enumerate}
    \item The cluster environment has dramatically quenched its member galaxies with a quiescent fraction of $88^{+4}_{-20}$ per cent within 0.5$r_{500}$, significantly enhanced relative to the field value of $20^{+2}_{-2}$ per cent at $z\sim2$.
    \item We find an excess of ``bulge-like'' quiescent cluster members with \Sersic index $n>2$ relative to the field but see no evidence for any physical segregation of these members within the cluster itself.
    \item At a fixed mass of $M_\star=5\times10^{10}M_\odot$, bulge-like, quiescent galaxies in the cluster are larger than their field counterparts at 99.6 per cent confidence. This suggests that these cluster member galaxies have experienced an accelerated size evolution relative to the field at $z>2$.
\end{enumerate}

XLSSC 122 is the first and only ICM selected cluster observed at $z\sim2$ and current observations only cover out to the cluster virial radius ($R_{vir}$). Extending observations of XLSSC 122 out to larger radii will allow for more detailed comparisons with the environmental quenching present in low-$z$ clusters, which has been observed to operate out to $2-3R_{vir}$ \citep[e.g.][]{Boselli2006, vonderLinden2010}. 
Additionally, new X-ray and SZ surveys, such as eROSITA, SPT-3G, and Advanced-ACT, to be followed in the mid/late 2020s by CMB-S4 and Athena, will find potentially hundreds of comparably massive, ICM selected clusters at $z\gtrsim2$. Combined with HST and JWST observations, these clusters will provide powerful statistical insight into the physical mechanisms behind the quenching and size evolution of cluster member galaxies in the early universe.

\section*{Acknowledgements}

This work is based on observations made with the NASA/ESA Hubble Space Telescope (GO 15267, 12177, and 12328), which is operated by the Association of Universities for Research in Astronomy, Inc., under NASA contract NAS5-26555.
E.N., R.E.A.C., S.W.A., and A.M. acknowledge support from NASA grant number HST-GO-15267.002-A.
E.N. and J.P.W. acknowledge support from the Natural Sciences and Engineering Research Council of Canada. This work was supported in part by the U.S. Department of Energy under contract number DE-AC02-76SF00515. The Cosmic Dawn Center is funded by the Danish National Research Foundation. This research made use of Astropy,\footnote{http://www.astropy.org} a community-developed core Python package for Astronomy \citep{astropy:2013, astropy:2018}.

{\it
\section*{Data Availability}
The data underlying this article are available in the article and in its online supplementary material.
}




\bibliographystyle{mnras}
\bibliography{bibtex} 



\appendix

\section{Robustness of size measurements} \label{app:test}

\begin{figure}
 \includegraphics[width=0.98\columnwidth]{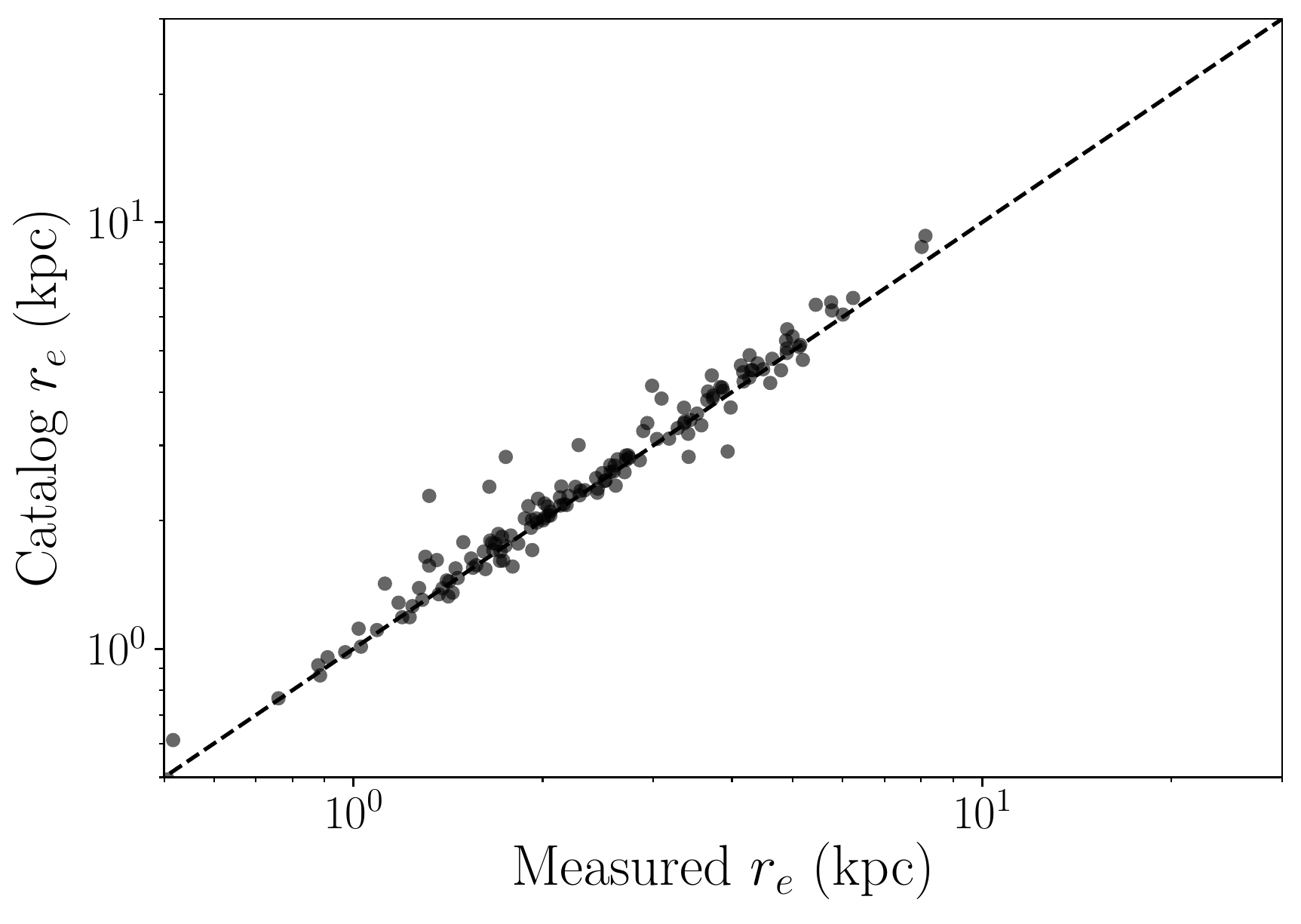}
 \caption{The effective radii of COSMOS sources measured using our fitting pipeline (see \cref{sec:galfit}) is compared to that cataloged by \protect\cite{VanDerWel2012}.
Sources are selected to best match those in our analysis (see \cref{sec:selection}) by filtering to $1.9<z<2.1$, F140W~$<24$, and $\log(M_\star/M_\odot )>10.3$. Size measurements are made in the F125W band. The mean offset between the two sets of measurements is 2.4 per cent.}
 \label{fig:cosmos_compare}
\end{figure}

We test the efficacy of our \verb+GALFIT+ pipeline in determining accurate half-light radii by refitting a subset of COSMOS galaxies analysed in \cite{VanDerWel2012} and comparing to the cataloged radii. We select galaxies with $1.9<z<2.1$, F140W$<24$, and $\log(M_\star/M_\odot )>10.3$ in order to best match those in our analysis. We perform the measurements in the F125W band as it is the closest band to F140W that has published structural parameters. Fitting is performed in an identical fashion to that described in \cref{sec:galfit} with the only difference being that we utilize the published F125W PSF from the \cite{VanDerWel2012} study.

The results of this test are shown in \cref{fig:cosmos_compare}. We find an average offset between our measurements and those cataloged in \cite{VanDerWel2012} of 2.4 per cent. Our measurements are on average smaller by this amount. This level of disagreement is well within expectations for the high-redshift, low signal-to-noise sources that we are focused on and is negligible for the purposes of our study. The mean cataloged error on the sizes of these sources in \cite{VanDerWel2012} is 6.4 per cent. The fitting region and convolution box-sizes described in \cref{sec:galfit} have been tuned to maximize the agreement of our measured radii to the cataloged radii in this test.

\section{XLSSC 122 cluster members}

\begin{table*}
 \caption{The properties of the 28 member galaxies of XLSSC 122 studied in this paper (see \cref{sec:selection}) are tabulated. Type references quiescent (Q) or SFG (S). Colours are expressed as F105W - F140W magnitudes measured in 0.8 arcsecond circular apertures. The F140W magnitude quoted employs a Kron-type aperature (see \protect\cite{Willis2020}). While not utilized in this study, the circularized effective radius ($r_e^{\mathrm{circ}}=r_e\sqrt{q}$) is included for posterity. Stellar masses for quiescent galaxies are derived in \protect\cite{Willis2020} and adjusted following \cref{sec:masses}. A digital version of this table, including errors on $n$ and $q$, is available in the online supplementary material.
}
 \label{tab:members}
\begin{tabular}{rrrrrrrrrrrr}
\toprule
ID & RA (deg) & Dec (deg) & Type & Colour & F140W & $\log M_\star/M_\odot$ & $r_e$ (kpc) &  $r_e^{\mathrm{circ}}$ (kpc) & $n$ & $q$ & $r/r_{500}$\\
\midrule
  529 &  34.4342 & -3.7588 &    Q &   1.44 &  20.636 &  $11.97\pm0.04$ &  $49.45\pm6.08$ &  $44.75\pm5.51$ &  7.56 &  0.82 &      0.01 \\
  455 &  34.4223 & -3.7635 &    Q &   1.29 &  21.950 &  $10.84\pm0.09$ &   $2.37\pm0.04$ &    $2.30\pm0.04$ &  2.94 &  0.94 &      1.31 \\
  661 &  34.4341 & -3.7577 &    Q &   1.49 &  21.668 &  $11.09\pm0.03$ &    $2.31\pm0.10$ &   $2.19\pm0.09$ &  3.07 &  0.90 &      0.11 \\
 1036 &  34.4324 & -3.7499 &    Q &   1.33 &  22.385 &   $10.86\pm0.10$ &   $3.92\pm0.47$ &   $2.81\pm0.34$ &  5.03 &  0.51 &      0.92 \\
  300 &  34.4350 & -3.7679 &    Q &   1.56 &  22.503 &  $10.99\pm0.01$ &   $1.14\pm0.02$ &   $0.98\pm0.02$ &  2.57 &  0.74 &      0.95 \\
  920 &  34.4356 & -3.7531 &    S &   0.43 &  22.725 &               - &   $0.75\pm0.02$ &   $0.69\pm0.02$ &  6.39 &  0.85 &      0.60 \\
  305 &  34.4472 & -3.7680 &    Q &   1.42 &  22.525 &  $11.02\pm0.08$ &   $5.82\pm0.48$ &   $3.73\pm0.31$ &  7.30 &  0.41 &      1.64 \\
 1057 &  34.4369 & -3.7502 &    Q &   1.37 &  22.848 &  $10.61\pm0.11$ &    $1.70\pm0.02$ &   $1.22\pm0.02$ &  1.63 &  0.51 &      0.93 \\
 1065 &  34.4359 & -3.7495 &    Q &   1.35 &  22.341 &  $10.56\pm0.11$ &   $2.35\pm0.16$ &   $1.43\pm0.13$ &  0.94 &  0.37 &      0.96 \\
  608 &  34.4385 & -3.7607 &    Q &   1.28 &  22.991 &  $10.35\pm0.12$ &   $1.27\pm0.02$ &   $1.09\pm0.02$ &  1.21 &  0.74 &      0.49 \\
  243 &  34.4224 & -3.7700 &    Q &   1.20 &  22.617 &  $10.26\pm0.14$ &   $1.21\pm0.02$ &   $0.94\pm0.02$ &  2.17 &  0.60 &      1.67 \\
  847 &  34.4347 & -3.7549 &    Q &   1.38 &  23.382 &  $10.34\pm0.15$ &    $0.90\pm0.02$ &   $0.83\pm0.02$ &  1.94 &  0.84 &      0.40 \\
  375 &  34.4441 & -3.7657 &    S &   0.66 &  23.082 &               - &    $2.66\pm0.10$ &    $2.46\pm0.10$ &  2.89 &  0.85 &      1.25 \\
  735 &  34.4250 & -3.7580 &    Q &   1.43 &  23.390 &  $10.52\pm0.11$ &   $1.41\pm0.04$ &   $0.98\pm0.03$ &  2.60 &  0.48 &      0.94 \\
 1223 &  34.4433 & -3.7450 &    Q &   1.31 &  23.494 &  $10.19\pm0.17$ &   $0.73\pm0.02$ &   $0.65\pm0.02$ &  2.19 &  0.78 &      1.70 \\
  347 &  34.4418 & -3.7667 &    Q &   1.27 &  23.561 &  $10.05\pm0.21$ &   $2.29\pm0.14$ &   $1.86\pm0.11$ &  2.06 &  0.66 &      1.14 \\
  146 &  34.4448 & -3.7729 &    S &   0.48 &  22.902 &               - &   $3.65\pm0.05$ &   $2.55\pm0.04$ &  0.90 &  0.49 &      1.82 \\
  497 &  34.4330 & -3.7632 &    Q &   1.38 &  23.579 &  $10.31\pm0.17$ &   $1.14\pm0.03$ &   $1.12\pm0.03$ &  2.44 &  0.98 &      0.47 \\
  604 &  34.4394 & -3.7603 &    S &   0.46 &  23.375 &               - &   $4.34\pm0.04$ &   $2.11\pm0.03$ &  0.37 &  0.24 &      0.56 \\
 1147 &  34.4336 & -3.7477 &    S &   0.47 &  23.847 &               - &   $0.86\pm0.02$ &   $0.67\pm0.02$ &  0.86 &  0.61 &      1.13 \\
  407 &  34.4464 & -3.7653 &    S &   0.66 &  23.046 &               - &   $2.57\pm0.22$ &    $1.80\pm0.16$ &  2.75 &  0.49 &      1.43 \\
  731 &  34.4398 & -3.7583 &    Q &   1.61 &  23.974 &   $10.36\pm0.10$ &   $0.76\pm0.03$ &   $0.63\pm0.02$ &  1.93 &  0.69 &      0.58 \\
  653 &  34.4340 & -3.7593 &    Q &   1.54 &  22.444 &   $10.32\pm0.10$ &     $1.40\pm0.10$ &   $1.18\pm0.09$ &  4.15 &  0.71 &      0.06 \\
  726 &  34.4306 & -3.7576 &    Q &   1.28 &  23.464 &  $10.11\pm0.28$ &   $5.37\pm0.16$ &   $4.52\pm0.14$ &  1.00 &  0.71 &      0.38 \\
  454 &  34.4190 & -3.7639 &    S &   0.45 &  23.689 &               - &   $3.71\pm0.07$ &   $2.88\pm0.06$ &  0.45 &  0.60 &      1.65 \\
  808 &  34.4477 & -3.7561 &    S &   1.02 &  23.783 &               - &   $3.69\pm0.08$ &   $2.26\pm0.05$ &  0.75 &  0.38 &      1.42 \\
  554 &  34.4353 & -3.7625 &    S &   0.28 &  23.960 &               - &   $2.53\pm0.04$ &   $1.29\pm0.03$ &  0.36 &  0.26 &      0.40 \\
  434 &  34.4466 & -3.7645 &    Q &   1.38 &  23.887 &  $10.18\pm0.27$ &   $6.22\pm0.67$ &   $3.93\pm0.43$ &  2.61 &  0.40 &      1.41 \\
\bottomrule
\end{tabular}
\end{table*}


\bsp	
\label{lastpage}
\end{document}